\documentclass[pre,showpacs]{revtex4}

\usepackage{graphicx}

\newcommand{\be}{\begin{equation}}  
\newcommand{\ee}{\end{equation}}  
\newcommand{\bi}{\begin{itemize}}  
\newcommand{\ei}{\end{itemize}}  
\newcommand{\bfg}{\begin{figure}[ht] \begin{center}}  
\newcommand{\efg}{\end{center} \end{figure}}

\begin{document}

\title{Learning and predicting time series by neural networks}

\author{Ansgar Freking}
\author{Wolfgang Kinzel}
\affiliation{Institut f\"ur Theoretische Physik und Astrophysik,
  Universit\"at W\"urzburg, Am Hubland, 97074 W\"urzburg, Germany}
\author{Ido Kanter}
\affiliation{Minerva Center and Department of Physics, Bar-Ilan
  University, Ramat-Gan 52900, Israel}

\begin{abstract}
  Artificial neural networks which are trained on a time series
  are supposed to achieve two abilities: firstly to predict the series
  many time steps ahead and secondly to learn the rule which has
  produced the series. It is shown that prediction and learning are not
  necessarily related to each other. Chaotic sequences can be learned
  but not predicted while quasiperiodic sequences can be well predicted
  but not learned.
\end{abstract}


\maketitle

Neural networks are able to learn a rule from a set of examples.  This
paradigm has been used to construct adaptive algorithms -- named
artificial neural networks  -- which are trained on a set of
input/output patterns generated by an unknown function. After the
training process, the network can reproduce the patterns, but is also
has achieved generalization: it has obtained some knowledge about the
unknown function. 

In the simplest case the unknown function is a neural network itself,
the ``teacher''. A different neural network with an identical
architecture, the ``student'', is trained on a set of examples produced
by the teacher.  This so called "student/teacher" scenario has been
intensively studied using models and methods of statistical physics
\cite{Hertz,EvdB,BC}.  Recently these methods have also been applied to
learning and generation of time series \cite{EKKK,KKPE,SK,EK,PK}.

The main result of these theoretical investigations is that as the
student network receives more information it increases its similarity to
the weights of the teacher network. When the number of training examples
is much larger than the number of parameters of the teacher, the student
is almost identical to the teacher and the generalization error is close
to zero.  In this article we show that this fundamental relation between
learning and generalization is violated when a neural network is trained
on a time series. We present a class of networks with almost perfect
prediction of the series and almost zero information about the rule.
The opposite case is found, as well: A network cannot predict a time
series although it is almost identical to the rule generating the
series.

Hence the intuitive deduction that learning a rule leads to good
generalization and good generalization indicates good knowledge about
the rule is violated both ways when a neural network is trained on a
time series.

We find this phenomenon already for a simple perceptron, a neural
network with a single layer of synaptic weights, given by the equation
$o = g \left( \mathbf{w} \cdot \mathbf{S} \right)$.  Here
$\mathbf{w}=(w_1,w_2,...,w_N)$ is the vector of synaptic weights,
$\mathbf{S}=(s_{t-1},s_{t-2},...,s_{t-N})$ is the input of the network
(window of the time series), $o$ is the output value and $N$ is the size
of the network. In the following we will study different transfer
functions $g(x)$. Such a perceptron can be used as a sequence generator
(teacher with weights $\mathbf{w}^T$) as well as a network being trained
on a time series (student with weights $\mathbf{w}^S$)\cite{EKKK}.

The sequence is generated by a teacher network with random weights,
starting from random initial conditions $(s_N,s_{N-1},...,s_1)$;
hence it is defined by the equation
\be s_t = g\left( \sum\limits_{j=1}^N w_j^T s_{t-j}\right) 
\label{drei}
\ee 
We define a time $t_0$ in such a way that the sequence is stationary for
any $t>t_0$. Here, "stationary" means that the sequence lies on its
attractor. The transient, which is of $\mathcal{O}(N)$ is not included
in the training examples.

The training error is calculated from the absolute value of the
deviation between the sequence $s_t$ and the corresponding output $o_t$
of the student:
\be
\epsilon=\lim\limits_{T\rightarrow\infty}\frac{1}{T}
\sum\limits_{t=t_0+1}^{t_0+T} |s_t-o_t|\label {zwei} 
\ee
This is the average error of a one-step-prediction of the student on the
time series.  Perfect training leads to zero error $\epsilon$, meaning
that each number of the sequence is correctly reproduced: $s_t=o_t$.

The student's knowledge about the unknown parameters is measured by the
overlap $R$ between the weight vectors of the teacher and the student:
\be R=\frac{ \mathbf{w}^T \cdot \mathbf{w}^S} {|\mathbf{w}^T|
  |\mathbf{w}^S|}
\label{vier} 
\ee
If the transfer function is continuous, it is also important that the two
vectors coincide in their length $Q^S=Q^T$ with $Q=|\mathbf{w}|$.

First we discuss the Boolean perceptron, $g(x)=sign(x)$, of size $N$
which has generated a periodic bit sequence \cite{EKKK,SK}. The teacher
perceptron has random weights with zero bias, and the cycle is related
to one component of the power spectrum of the weights. The student
network is trained using the perceptron learning rule:
\begin{eqnarray}
\label{dreizehn}
\Delta w_i^S & =  \frac{1}{N} s_t \; s_{t-i} & \; \;\mbox{if} \; \; s_t 
\sum\limits^{N}_{j=1} w_j^S s_{t-j} <0; \nonumber \\ 
\Delta w_i^S & = 0 \mbox{~~else}.
\end{eqnarray}
For this algorithm there exists a mathematical theorem
\cite{Hertz}: If the set of examples can be generated by
some perceptron then this algorithm stops, i.e.  it finds one out of
possibly many solutions.  Since we consider examples from a bit
sequence generated by a perceptron, this algorithm is guaranteed to
learn the sequence perfectly.

The network is trained on the cycle until the training error is zero.
Hence the student network can predict the stationary sequence perfectly.
It turns out that the overlap between student and teacher remains small,
in fact it is zero for infinitely large networks, $N\rightarrow \infty$.
Although the network predicts the sequence perfectly, it does not gain
much information on the parameters of the network which has generated
this sequence.

\bfg
\includegraphics[width=.5\textwidth]{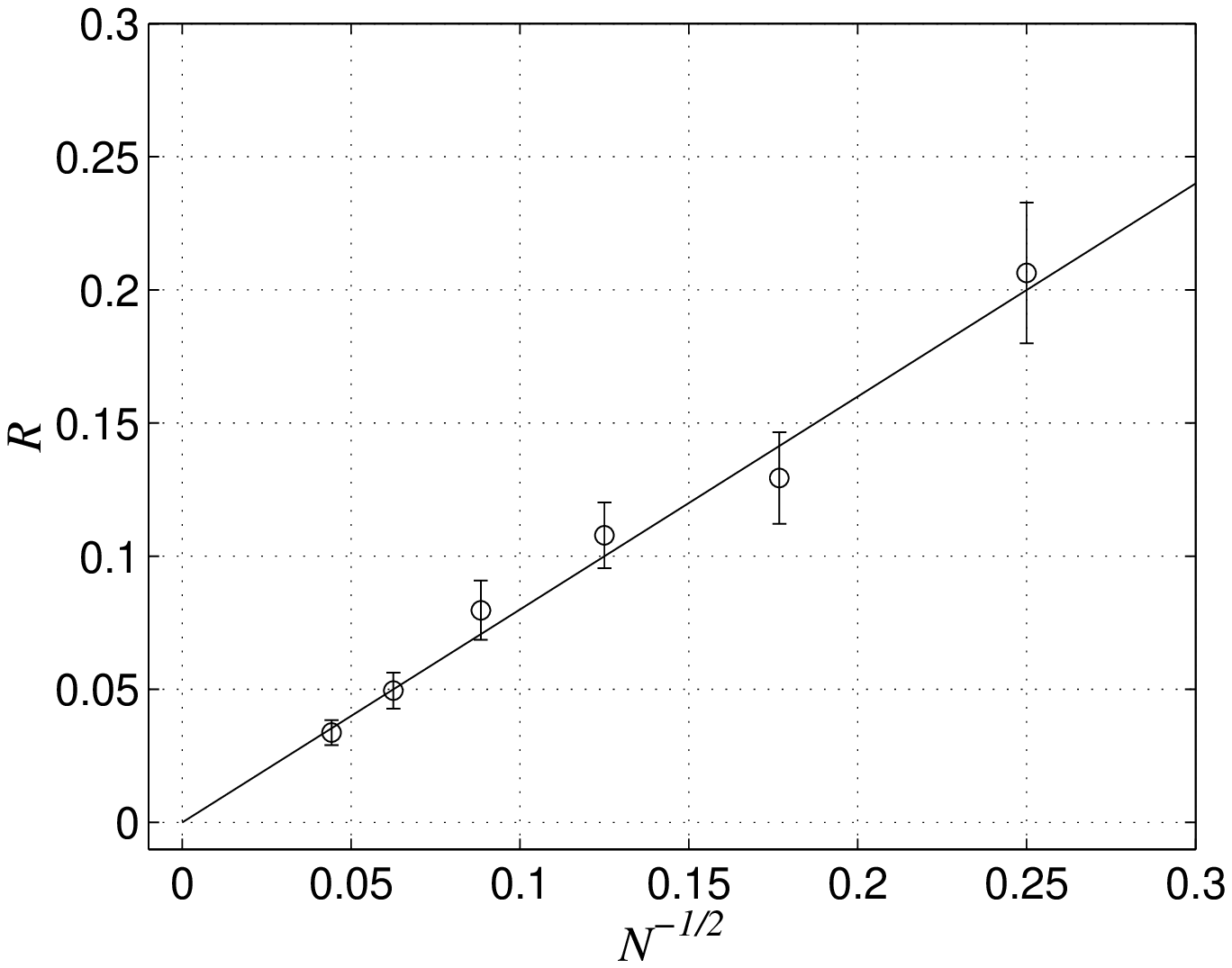}
\caption{Final overlap $R$ between student and teacher network 
after training, as a function of the size $N$ of the network. 
The standard error-bars result from $M=100$ individual runs.
A linear fit of $R$ vs.\ $N^{-1/2}$ supports the statement that
$R\rightarrow 0$ for $N\rightarrow \infty$.} 
\efg

This situation seems to be different in the case of a continuous
perceptron. Inverting Eq.~(\ref{drei}) for a monotonic transfer function
$g(x)$ gives $N$ linear equations for $N$ unknowns $w_i^T$.  If all
patterns are linearly independent then batch training, using $N$
windows, leads to perfect learning.

\bfg
\hfill
\includegraphics[width=.4\textwidth]{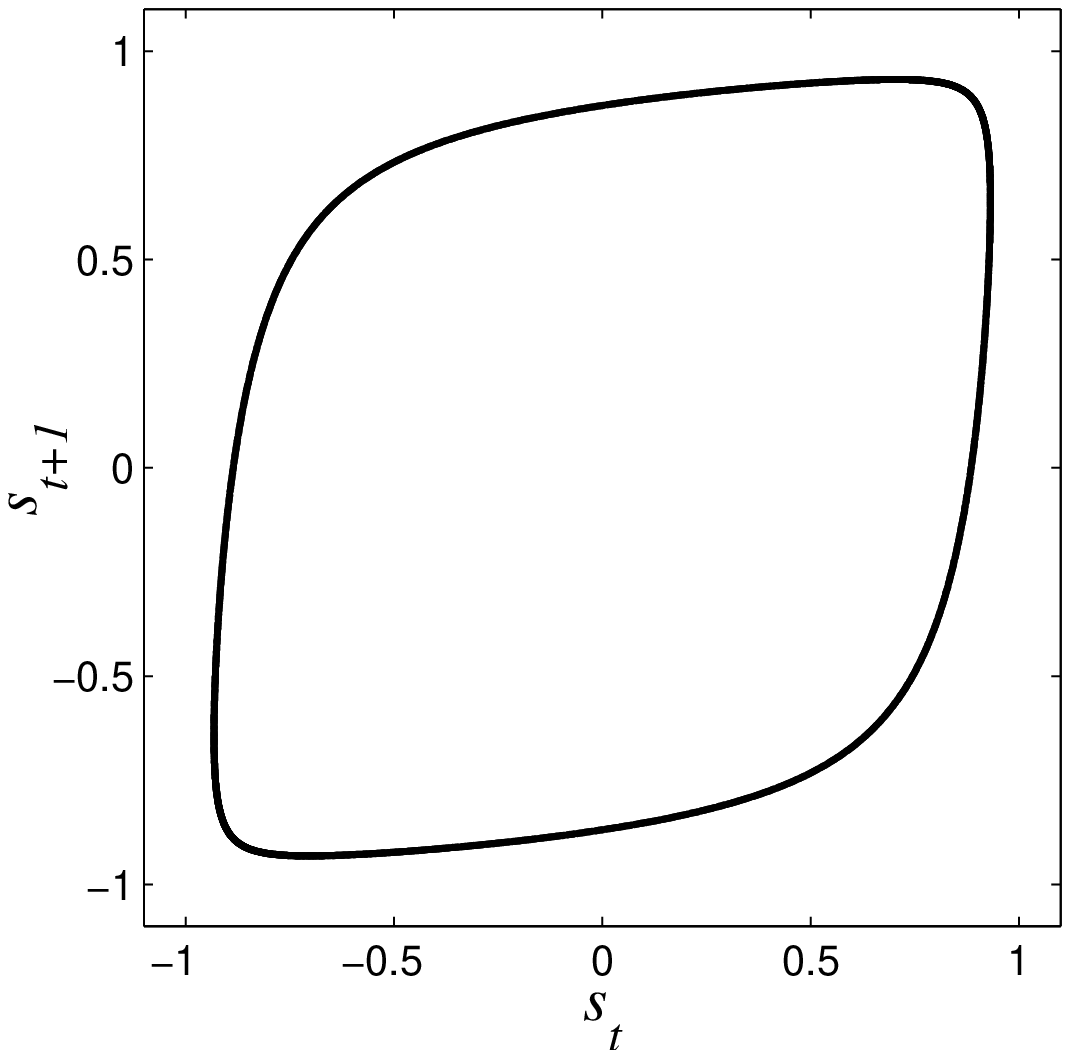} \hfill\hfill
\includegraphics[width=.4\textwidth]{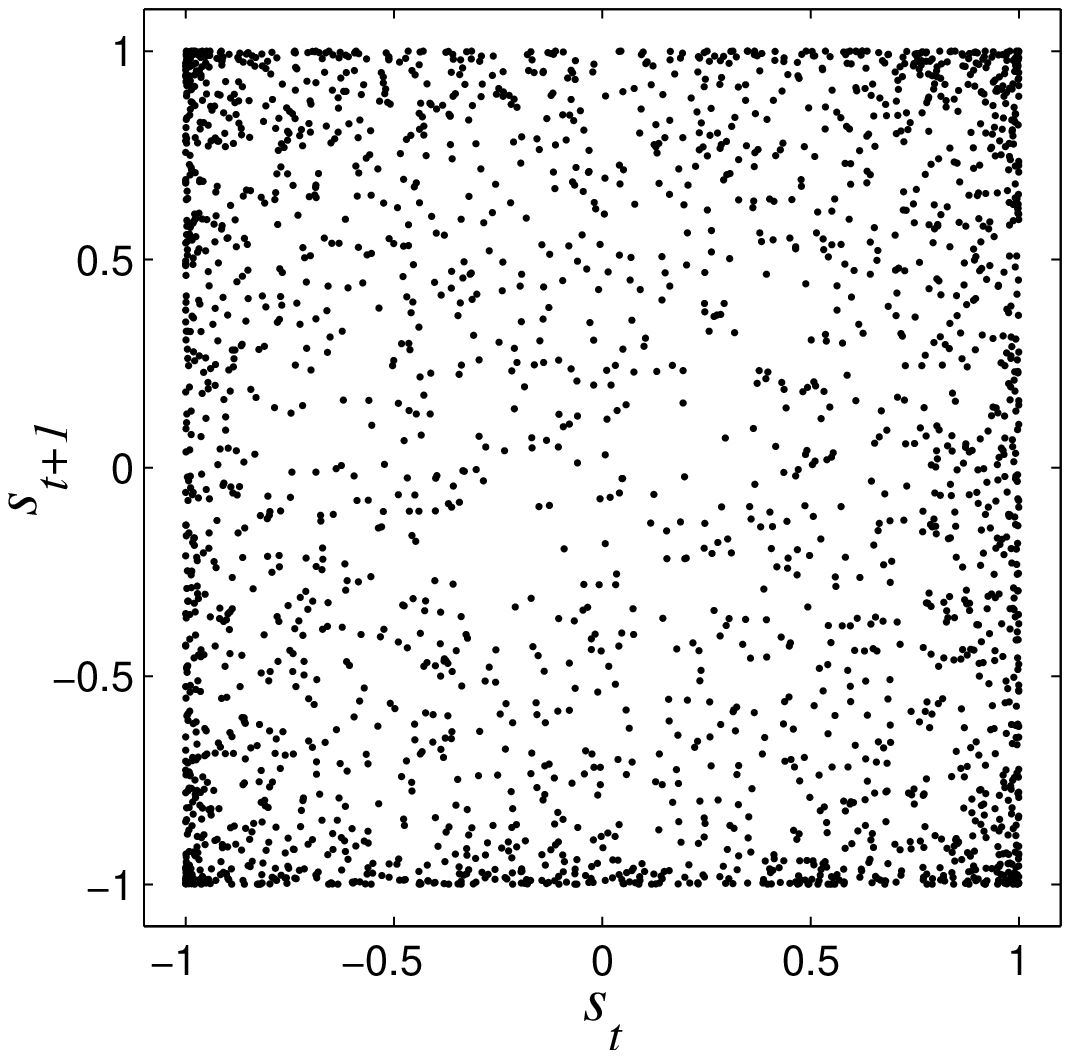} \hfill ~
\caption{\label{rmap_ts}
Return map for a quasiperiodic (left) and chaotic(right) 
time series used for training a perceptron as described in the text}
\efg

A network with transfer function $g(x)=\tanh(\beta\, x)$ generates a
quasiperiodic time series, if the parameter $\beta$ is larger than a
critical value $\beta_c$ \cite{KKPE}. 
The form of the sequence is characterized by an attractor of dimension
equal to one and analytically in the leading order it is given by
\be
s_t=\tanh(A \cos (2\pi\,q\,t/N)), 
\label{sieben}
\ee
with some gain $A(\beta)$, which is non-zero above the bifurcation point
$\beta_c$. Note, that in the typical sequence there's only a
contribution of one non-integer wavenumber $q$, which is related to one
dominant Fourier component of the couplings $\mathbf{w}^T$, see
\cite{KKPE} for details.

When trying to find the couplings $\mathbf{w}^T$ by inverting the set
of Eq.~\ref{drei}, it turns out that even professional computer
routines often fail to perform the required matrix inversion: the
patterns are almost linearly dependent. Some explanation for that can
be found from Eq.~\ref{sieben}. 
For small $A$, the tanh in Eq.~\ref{sieben} can be approximated by its
argument and one can easily show that $s_{t+2}=-s_t+2\cos(2\pi
q/N)s_{t+1}$. Therefore, any window of the sequence can be written as a
linear combination of two basis vectors. In case we expand the tanh in
Eq.~\ref{sieben} up to the $\rho$'s term, one can show that the form of
$s_{t+m}$ is given by $s_{t+m}(\rho)= \sum_{k=0}^{\rho-1} B_{2k+1} (
\cos(2\pi q/N(2k+1))-\sin(2\pi q/N(2k+1)))$, since $\cos(x)^{2\rho+1}
=\sum C_{2k+1} \cos((2k+1)x)$ where $C_k$ and $B_k$ are constants. On
one hand, as long as $\rho$ is less than the window size $N$, the inputs
are linearly dependent and Eq.~\ref{sieben} cannot be inverted. On the
other hand, the power expansion of the tanh indicates that $B_{\rho}$
drops exponentialy with $\rho$. Thus, the linear dependence of the
$N$-dimensional inputs is lifted only by the $\rho=N+1$ term in the
expansion which decreases exponentially as $N$ increases.  This is the
source for the ill-conditioned problem of inverting Eq.~\ref{sieben}.

Hence, in particular for large dimensions $N$, batch learning does
not work well for quasiperiodic time series generated by a teacher
perceptron.

How does this scenario show up in an on--line training algorithm for
a continuous perceptron? If a quasiperiodic sequence is learned step
by step using gradient descent to update the weights, without
iterating previous steps,
\begin{equation}
\label{vierzehn}
\Delta w_i^S = \frac{\eta}{N} (s_t - g(h)) \cdot g^{\prime} (h) \cdot
s_{t-i} \; \; \mbox{with} \; \; h= \beta \sum\limits^{N}_{j=1} w_j^S s_{t-j}
\end{equation} 
we find two time scales (time = number of training steps):
(i) A fast one increasing the overlap between teacher and student to
  a value which is still far away from  perfect agreement, $R=1$ and
  $Q^S=Q^T$.
  During this phase, the training error goes down to nearly zero.
(ii) A slow one further increasing the overlap and still decreasing the
  training error.
  
  Since the second time scale is usually several orders of magnitude
  larger than the first one, we could not observe $R=1$ within our
  numerical simulations.  Although there is a mathematical theorem on
  stochastic optimization which seems to guarantee convergence to zero
  training error (\ref{zwei}) \cite{StochOpt}, which implies full
  overlap $R=1$ with $Q^S=Q^T$, our on--line algorithm cannot gain much
  information about the teacher network, at least within practical
  times.

\bfg
\includegraphics[width=.6\textwidth]{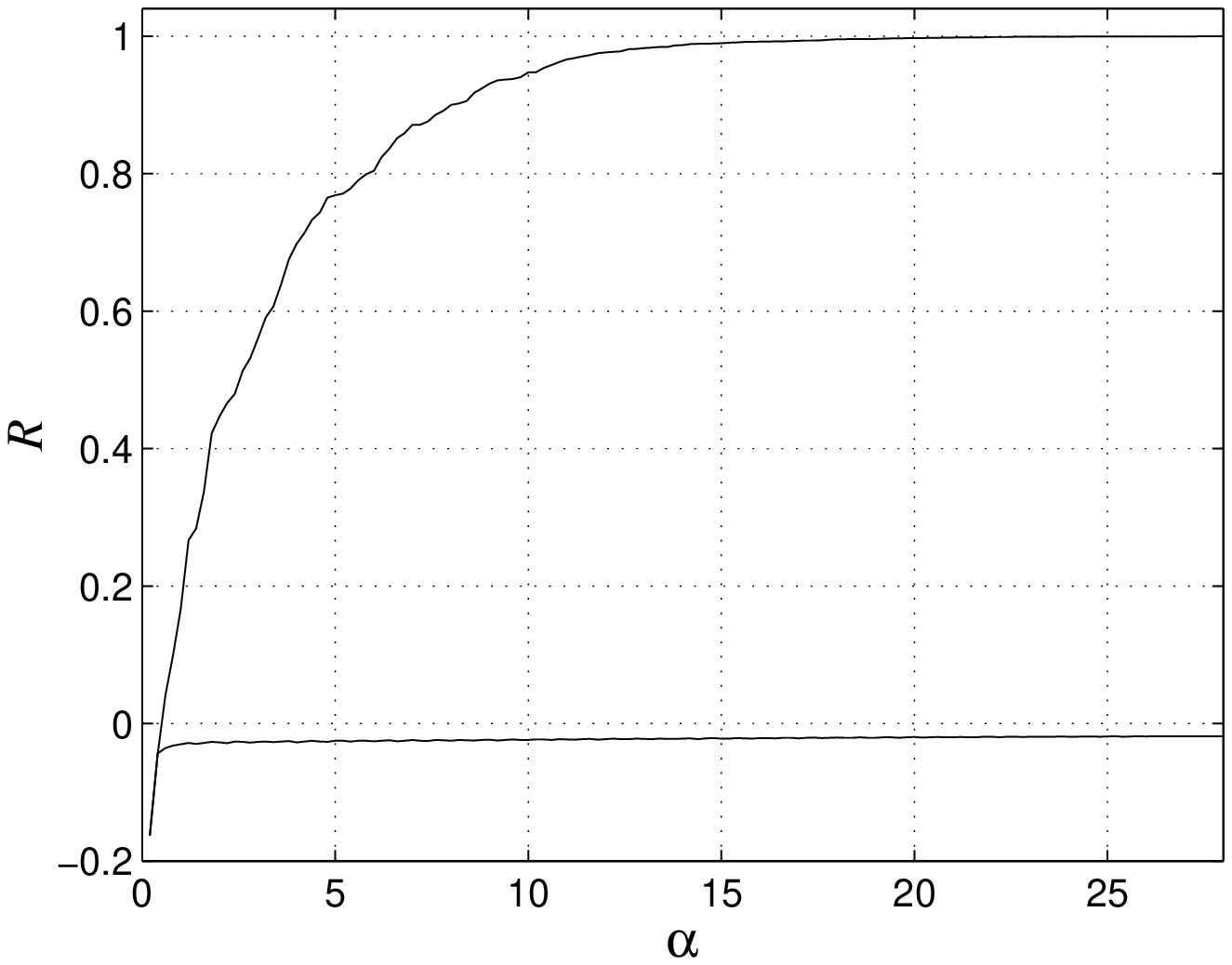}
\caption{Overlap $R$ as a function of the fraction $\alpha=t/N$ of training
  examples. 
  The upper curve shows the learning dynamic for the chaotic case, the
  lower one shows the two time scales for the training on the
  quasiperiodic series. Both settings start with the same initial
  overlap ($R_0\approx -0.16$).  At $\alpha \approx 0.5$ the dynamics of
  the quasiperiodic case enters the part with slow progress.}
\label{rdyn} 
\efg

This is completely different for a chaotic time series generated by a
corresponding teacher network with $g(x)=\sin(\beta\, x)$ \cite{PK}. It
turns out that learning the chaotic series works like learning random
examples: After a number of training steps of the order of $N$ the
overlap $R$ relaxes exponentially fast to perfect agreement between
teacher and student, $R=1$.  The same behavior can be observed for the
length $Q^S$ of the student, which approaches exponentially fast to the
length of the teacher.

Here are some details of the numerical calculations:
Our simulations were performed with the same (random) teacher
weights for the quasiperiodic and the chaotic case. Furthermore, the
random initialization of the student networks were identical.  The
settings differ only in the choice of the transfer-functions $g(x)$.
Return maps for the two sequences are shown in Fig.~\ref{rmap_ts}.

Starting with the same initial overlap, the students were trained
according to Eq.~(\ref{vierzehn}) until they achieved a certain training
error ($\epsilon=0.008$). In both cases this took about $25 N$ learning
steps, the network dimension was $N=50$.  After the training process
however, the students ended up with completely different weight vectors.
In case of the chaotic sequence, the student's weights came close to the
one of the teacher ($R\rightarrow 1$,$Q^S\rightarrow Q^T$).  In
contrast, the student of the quasiperiodic sequence did not obtain much
information about the teacher, and its weights remained nearly
perpendicular to the teacher ones ($R\approx 0$).  The time evolution of
the respective overlaps during training is shown in Fig.~\ref{rdyn}.

One important question remains: How well can the student predict the
time series? In order to evaluate the training success, we have defined
a \textit{one}-step-error in Eq.~(\ref{zwei}). Now we are interested in
the long-term prediction of the students. Therefore, the student
perceptrons have to act as sequence generators themselves, using their
own output to complete the next input window. Starting from a window of
the teacher's sequence, i.e.\ 
$(o_t,o_{t-1},...o_{t-N+1})=(s_t,s_{t-1},...s_{t-N+1})$,
the student's prediction $\tau$ steps ahead is given by iterating
Eq.~(\ref{drei}) up to
\be
o_{t+\tau}=g\left( \beta \sum_{j=1}^N w_j^S o_{t+\tau-j} \right) 
\label{fuenf}
\ee
The prediction error $\epsilon(\tau)$ is the average absolute deviation
of this value with the respective item of the teacher's sequence,
\be
\epsilon(\tau)=\lim\limits_{T\rightarrow\infty}\frac{1}{T}
\sum\limits_{t=t_0+1}^{t_0+T} |s_{t+\tau}-o_{t+\tau}|\label{sechs}
\ee
Note, that the average is performed by changing the initial time
window. Again, $t_0$ is used to indicate any time step of the stationary 
part of the sequence.
To calculate $\epsilon(\tau)$ in the simulations, we have chosen
$T=N=50$. The result is shown in Fig.~\ref{figpred}.

\bfg
\includegraphics[width=.6\textwidth]{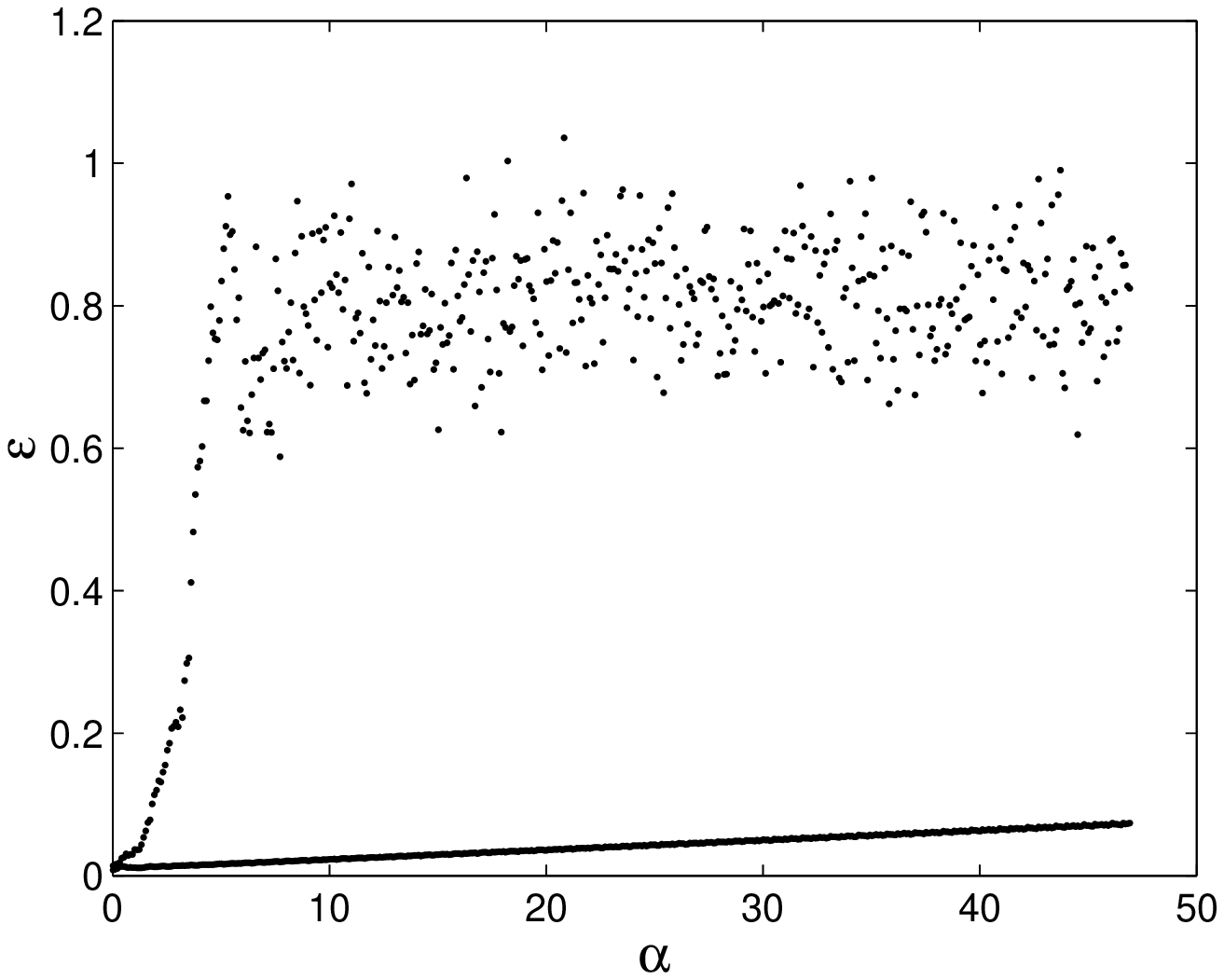}
\caption[]{Prediction error as a function of 
  time steps ahead (measured in multiples of $N$: $\alpha=\tau/N$), for
  the quasiperiodic (lower) and the chaotic (upper) series.}
\label{figpred}
\efg

The graph shows the prediction error as a function of the
time interval over which the student makes predictions. Both curves
coincide in the first value, which is equal to the
training error at which learning was stopped. 

The student network which has been trained on the quasiperiodic sequence
can predict it very well. The error increases linearly with the size of
the interval, even predicting $25N$ steps ahead yields an error of less
than 5\% of the total possible range. On the other side, the student
trained on the chaotic sequence cannot make long-term predictions. The
prediction error increases exponentially with time until it is of the
order of random guessing.

Of course, if the student would reproduce the series perfectly, it would
also predict it without errors. But since we stop our algorithm when the
training error is close but not identical to zero, we achieve two
different states: For the quasiperiodic sequence the weight vector of
the student recovers the main Fourier component of the teacher which
reproduces the sequences reasonably well. There remains a large space of
weight vectors which can generate the same sequence. For the chaotic
sequence, however, all the weights of the students come extremely close
to the ones of the teacher; but due to sensitivity to model parameters,
any prediction of the sequence is impossible.

All of our results stem from numerical simulations. We find that the
quantitative details of our results strongly depend on the parameters of
our model. Hence we did not succeed to derive quantitative results about
scaling of learning times with system size $N$ or the Ljuapunov exponent
as a function of the fractal dimension of the chaotic time series.

In summary we obtain the following result:

(i) A network trained on a quasiperiodic sequence does not 
obtain much information  about the teacher network which generated 
the sequence. But the network can predict this sequence 
over many (of the order of $N$) steps ahead.
(ii) A network trained on a chaotic sequence, however,  obtains almost 
complete knowledge about the teacher network. But due to the chaotic
nature of the sequence, this network cannot make reasonable predictions.

\enlargethispage{2cm}

\end{document}